# The influence of nonradiative relaxation on laser induced white emission properties in Cr:YAG nanopowders

M. Chaika, R. Tomala, O. Bezkrovnyi, W. Strek

Institute of Low Temperature and Structure Research, Polish Academy of Sciences, Wroclaw, Poland

**Abstract**

Laser Induced White Emission (LIWE) is the subject of research worldwide. Since its discovery, the understanding of this phenomenon has progressed successfully enough to reach industrial applications. However, a lack of understanding of the nature of this phenomenon limits its potential. This article is devoted to the study of the influence of nonradiative relaxation processes on the properties of laser induced white emission in Cr:YAG nanopowders. The concentration series of Cr:YAG nanopowders was synthetized by Pechini method. The microstructure, optical and LIWE properties were studied. The influence of chromium concentration on the number of photons involved in LIWE process (N parameter) is shown. The increase of N parameter is associated with an increase in the probability of non-radiative recombination processes with an increase of chromium concentration. A multiphoton ionization model is proposed to describe LIWE phenomenon.

## 1. Introduction

Since Tanner's discovery of Laser Induced White Emission (LIWE) in 2010, it was widely used for practical applications, but its potential remained limited by the lack of understanding of the nature of this phenomenon. Excitation of a sample by laser beam with energy density above the threshold in vacuum leads to the appearance of white light emission covering entire visible and near-infrared regions. Remarkably, the broadband spectra can be generated using a wide range of laser sources from blue to NIR laser diodes [1–3]. The main advantage of these materials in comparison to traditional LEDs based on YAG:Ce is the ability to generate continuous spectra with a high color rendering index making it possible to use them for obtaining artificial light [1,4,5]. Progress in understanding LIWE phenomenon is already sufficient for application of this effect for hydrogen generation [6]. However, despite the progress achieved in recent decades, the intensity of white light emission usually is relatively low.

A number of LIWE explanations based on the properties of the investigated materials were proposed. These explanations include black body radiation, photon avalanche, thermal avalanche, electron-hole recombination, intervalence charge transfer, etc. [7]. LIWE can be generated in various materials from graphene to organometallic compounds with similar white light emission properties meaning that the mechanism must be the same for all these materials. It has recently been suggested that the multiphoton ionization may be responsible for LIWE phenomenon [2,8,9]. This assumption is based on our recent findings that LIWE has a surface origin. The white light emission can only be emitted from the surface of the sample and do not penetrate to the bulk [8]. This theory has been successfully applied to explain LIWE features such as the pulsation of white light emission [2]. However, this theory is far from perfect and a lot of questions remain unclear.

One of the drawbacks of the multiphoton ionization theory is the lack of explanation of N parameter dependence on the concentration of doping impurities. The multiphoton ionization concept was previously used to explain the interaction of femtosecond laser pulses with atoms. Similar to LIWE, multiphoton ionization exhibits threshold behavior, exponential growth of emission intensity, and saturation of emission intensity [2]. Multiphoton ionization was proposed by Keldysh [10] in 1965 as an explanation of the effect of simultaneous absorption of photons with a total energy higher than the ionization potential. In accordance to this theory, the number of photons absorbed to produce LIWE (denoted as N) should be relatively close for the same materials independent on the doping concentration. However, some authors report a significant change of N parameter with a change of the doping concentration [7,11]. For example, $Y_{2(1-x)}Nd_{2x}Si_2O_7$ nanocrystals show an increase in the N parameter from 2 for undoped to 6 for fully concentrated sample [11]. At present, the origin of this difference remains unclear.

The purpose of this work is to study the influence of $Cr^{3+}$ concentration on the properties of LIWE for Cr:YAG nanopowders. A concentration series of Cr:YAG nanopowders was synthetized. The effect of nonradiative relaxation on laser induced white emission properties in Cr:YAG nanopowders was investigated. The multiphoton ionization theory was used to explain the results of this work.

## 2. Experimental

The concentration series of Cr:YAG nanopowders was synthetized by Pechini method. Seven samples with different concentrations of Cr ions (0%, 0.1%, 0.5%, 1%, 3%, 10%, and 30% relative to aluminum ions) were prepared and labeled as Cr0, Cr0.1, Cr0.5, Cr1, Cr3, Cr10, and Cr30, respectively.

The microstructure, optical and LIWE properties of synthetized Cr,Yb:YAG nanopowders were investigated. The X-ray diffraction patterns were collected using Panalytical X'Pert pro X-ray powder diffractometer. The synthetized samples were analyzed by transmission electron microscopy using the Philips CM-20 SuperTwin microscope. Absorbance spectra were measured by Varian 5E UV-VIS-NIR spectrophotometer. Excitation, emission spectra, and lifetime were obtained at room temperature using an Edinburgh Instruments FLS980 fluorescence spectrometer. 975 nm laser was used as an excitation source, LIWE spectra were collected in vacuum using AVS-USB2000 Avantes and Ocean Optics NIRQuest512-2.5 spectrometers for the anti-Stokes and Stokes parts of the spectra. One can find a detailed explanation of the experimental part in the supplementary files.

## 3. Result and Discussion

### 3.1 Microstructure

X-ray structural analysis shows a pure garnet phase with an average grain size of several tens of nanometers. To confirm the purity of the measured samples, the X-ray diffraction patterns were collected for all samples (Fig. S1). The microstructure was calculated using Rietveld refinement analysis. Fig 1 shows an example of Rietveld analysis of XRD pattern of Cr1 sample. The analysis confirms the presence of a cubic yttrium aluminum garnet phase (Ia3d) without any other impurity phases. The calculated lattice parameters were from 12.01 Å to 12.08 Å with an average grain size of 15 nm to 35 nm. The calculated lattice parameters, average grain sizes and microstrain are collected in Table 1. There are no regularities in microstructure parameters depending on the chromium content. The calculated parameters varied in a certain range for different concentrations of chromium ions.

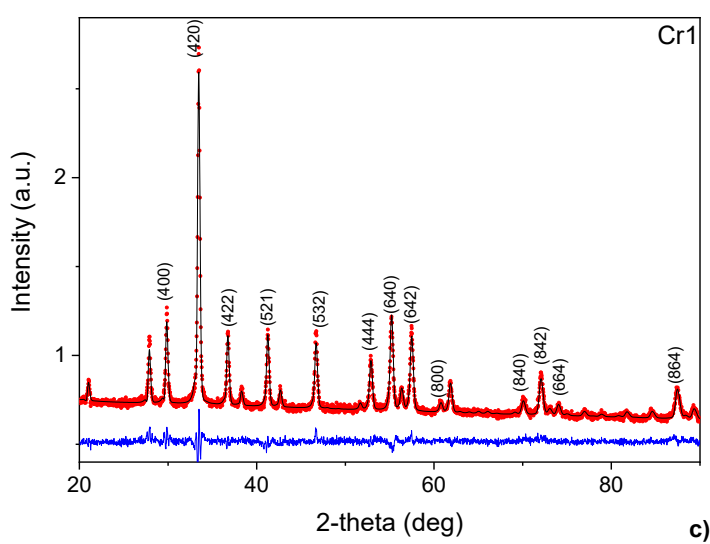

Fig. 1: X-ray diffraction pattern of the 1 at.% doped Cr:YAG nanopowders (red dots), results of Rietveld refinement analysis (black and blue curve).

Table 1: Cell parameter a, crystallite size D$_{XRD}$, and lattice microstrain ε for Cr:YAG nanopowders obtained by applying the Rietveld method to the XRD patterns.

| Sample | a, Å | D$_{XRD}$, nm | ε |
|---|---|---|---|
| Cr0 | 12,0125(6) | 25(1) | 0,0022(2) |
| Cr0.1 | 12,0240(4) | 35(1) | 0,0016(1) |
| Cr0.5 | 12,0312(5) | 34(1) | 0,0016(1) |
| Cr1 | 12,0073(5) | 22(1) | 0,0026(3) |
| Cr3 | 12,0412(5) | 36(1) | 0,0016(1) |
| Cr10 | 12,0818(9) | 18(1) | 0,0032(4) |
| Cr30 | 12,0412(5) | 32(1) | 0,0017(1) |

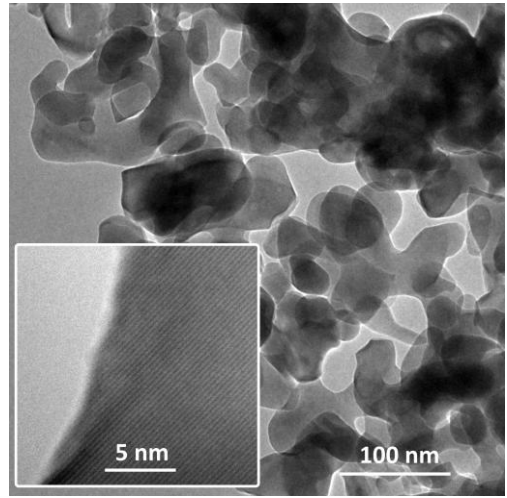

Fig. 2: TEM image of the Cr:YAG nanocrystals.

TEM study shows a fine crystal structure of the synthesized Cr:YAG nanocrystals confirming the results of the XRD studies. TEM image of the Cr5 sample is shown in the Fig. 2. The studied samples are characterized by fine crystalline structure, no traces of amorphous phases were found (inset on Fig. 2). All studied samples have the same morphology with the average particle size within several tens of nanometers in consistency with the XRD data. Our previous studies have shown that the addition of chromium, contrary to other dopants, changes the route of synthesis of YAG materials [12]. However, the synthesized Cr:YAG nanopowders have similar parameters, so it can be concluded that this difference has a significant influence on the spectroscopic properties. It should be noted that the earlier studies of YAG nanopowders doped with other dopants reveal clear dependence of the microstructure parameters on the concentration of dopants [13,14].

## 3.2 Optical properties

Absorption spectra of the Cr:YAG concentration series collected in the range of 200-800 nm are shown in Fig. 3(b). The samples are characterized by the presence of five main absorption peaks centered at ~210 nm, 270 nm, 370 nm, 450 nm and 600 nm corresponding to absorption of color centers, $Cr^{3+}$, and $Cr^{6+}$ ions. Two absorption bands centered at 450 nm and 600 nm correspond to $^4A_{2g}\rightarrow{}^4T_{2g}$ and $^4A_{2g}\rightarrow{}^4T_{2g}$ transitions of $Cr^{3+}$ ions [12,15–19]. An increase in chromium concentration leads to the rise of $Cr^{3+}$ absorption intensity. The strong absorption bands centered at 270 nm and 370 nm correspond to charge transfer transitions from oxygen to $Cr^{6+}$ ions [20]. Increasing the concentration of chromium did not affect the $Cr^{6+}$ absorption bands indicating only a small fraction of chromium ions in the 6+ valent states. An increase in the concentration of $Cr^{3+}$ ions leads to an increase in 210 nm absorption, which is most likely caused by an increase in the amount of defects (first of all, oxygen vacancies) [21]. However, the origin of the absorption centers at ~210 nm remains unclear and can be related as well to the band gap transitions (VB→CB).

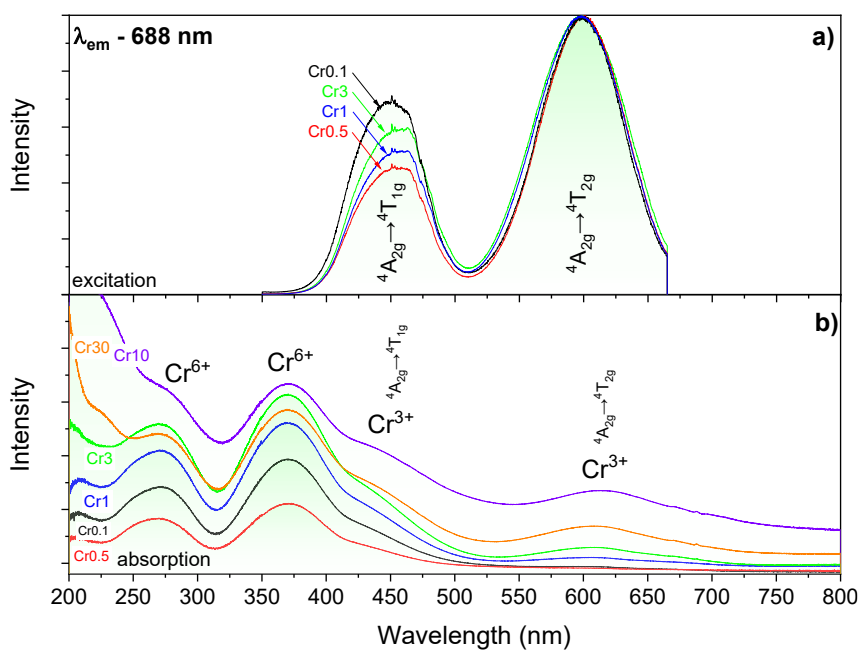

Fig. 3: Excitation (a) and absorption (b) spectra of Cr:YAG nanopowders.

Excitation spectra did not reveal energy transfer between $Cr^{6+}$ and $Cr^{3+}$ ions. The excitation spectra of Cr:YAG nanopowders were recorded at $\lambda_{em}$ = 688 nm, which corresponds to both $^4T_{2g}\rightarrow{}^4A_{2g}$ and $^2E_g\rightarrow{}^4A_{2g}$ transitions of $Cr^{3+}$ ions [22,23]. Two broad bands in the excitation spectra correspond to $^4A_{2g}\rightarrow{}^4T_{2g}$ and $^4A_{2g}\rightarrow{}^4T_{2g}$ transitions of $Cr^{3+}$ ions with no bands related to $Cr^{6+}$ ions. So, there is no charge transfer between $Cr^{6+}$ and $Cr^{3+}$ ions or, at least, it is weak. It should be noted that the excitation spectra of the Cr10 and Cr30 samples are not shown here due to low $Cr^{3+}$ emission intensity.

Cr$^{3+}$ luminescence increases with increasing the concentration up to 1 at.% with following reduction in the emission intensity. Photoluminescence spectra of the Cr:YAG nanopowders are shown in Fig. 4 (a). Excitation of the samples by 445 nm laser leads to appearance of strong red emission caused by both $^4T_{2g}\rightarrow{}^4A_{2g}$ and $^2E_g\rightarrow{}^4A_{2g}$ transitions [24]. The increase in the chromium doping content has no influence on the shape of Cr$^{3+}$ emission spectra. An integrated emission intensity increases with an increase in chromium concentration up to 1 at.% with the following decrease in emission intensity with increase in chromium concentration up to 30% (Table 2).

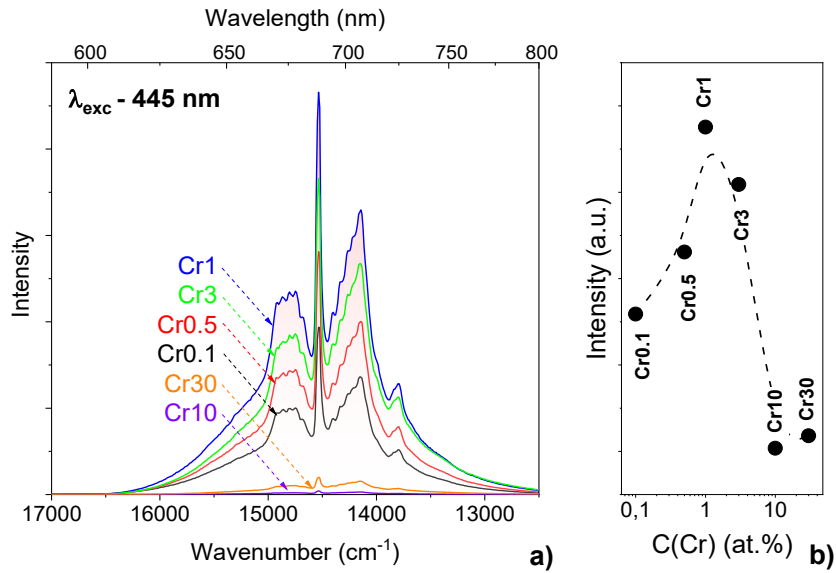

Fig. 4: Emission spectra (a) and PL intensity (b) of Cr$^{3+}$-doped YAG nanopowders measured at 445 nm excitation.

The fluorescence decay curves of Cr$^{3+}$ ions include two components that probably originate from Cr$^{3+}$ ions in different lattice positions. Fig. 5(a) shows the decay curves of R-lines ($\lambda_{em}$ = 688 nm) at $\lambda_{exc}$ = 450 nm for Cr:YAG nanopowders with chromium concentrations from 0.1 at.% to 30%. PL lifetime calculated from the decay curves using the double exponential approximation, is shown in Fig. 5(b). The lifetime of the slow decay component is in the range from 0.8 to 5.1 ms (blue dots in Fig. 5(b)), and the lifetime of the fast component is from 0.02 to 0.7 ms (red dots in Fig. 5(b)) (Table 2). The measured decay curves are the sum of both $^2E_g\rightarrow{}^4A_{2g}$ R-line and $^4T_{2g}\rightarrow{}^4A_{2g}$ broadband emission. It was previously found that the radiative decay curves of the R-lines can be fitted by a single exponential dependence, while in the case of the broadband transition the radiative decay curves consist from two components, which were detected in our case (see Fig. 5). The fastest component of the decay curves can be related to the Cr$^{3+}$ ions occupying the dodecahedral sites [25].

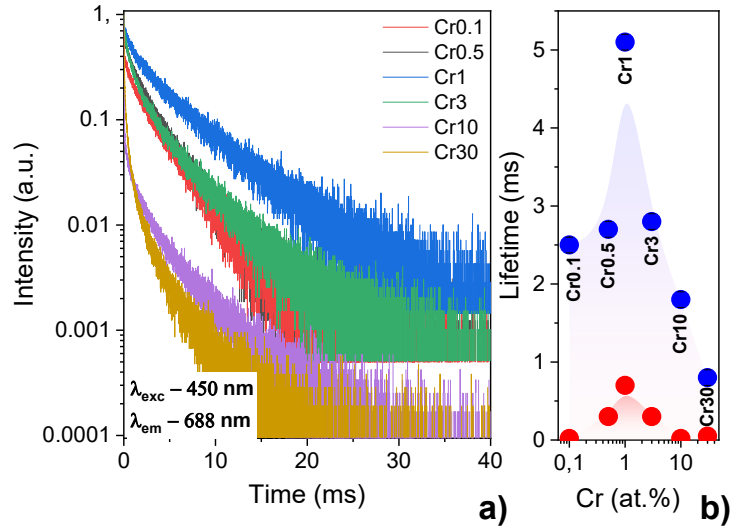

Fig. 5: Decay curves (a) and PL lifetime change (b) for the concentration series of Cr:YAG nanocrystals.

The measured lifetimes were almost the same for low concentrations of $Cr^{3+}$ ions (from 0.1 to 3 at.% with an exception of Cr1 sample). A further increase in the concentration of chromium ions leads to decrease of the lifetime (Table 2). We suppose that longer lifetime for Cr1 sample can be due to the higher crystal field strength compared to other samples. It was shown earlier that the process of Cr:YAG sintering differs from the sintering of pure YAG, for details see our previous paper [12]. This fact explains the lack of regularities in the lattice parameters for the series (Table 1) which can be clearly observed for other dopants [13].

Table 2: Emission intensity and lifetimes of $Cr^{3+}$ ions in Cr:YAG nanocrystals: $Cr^{3+}$ emission intensity at 450 nm ($I_{450}$), lifetime of the fast component of $Cr^{3+}$ ions ($\tau_1$) and lifetime of the slow component of $Cr^{3+}$ ions ($\tau_2$) measured at 450 nm excitation.

| Denote | $Cr^{3+}$ ($\tau_1$), ms | $Cr^{3+}$ ($\tau_2$), ms | $Cr^{3+}$ ($I_{450}$) |
|---|---|---|---|
| Cr0.1 | 2.52(1) | 0.022(1) | 42 % |
| Cr0.5 | 2.72(1) | 0.31(1) | 61 % |
| Cr1 | 5.14(2) | 0.68(1) | 100 % |
| Cr3 | 2.87(1) | 0.33(1) | 82 % |
| Cr10 | 1.78(1) | 0.026(1) | 0.9 % |
| Cr30 | 0.81(1) | 0.053(1) | 5 % |

## 3.3 LIWE properties

Under focused NIR laser beam the Cr:YAG nanopowders emit white light covering entire Vis and NIR regions from 400 nm to 2600 nm. The emission intensity in the visible part of the emission spectra increases with an increase of the emission wavelength. A similar pattern was found for NIR part of the emission spectra. It should be noted that these spectra were detected by CCD cameras, so the measured emission intensity is affected by the sensitivity of the camera. Previously, LIWE properties were investigated in Cr:YAG transparent ceramics [8,9,26]. The LIWE spectra of Cr:YAG transparent ceramics are characterized by weak emission intensity in the NIR region, which is a general trend for transparent materials [2,8,9,26]. At the moment, there is no explanation for this behavior.

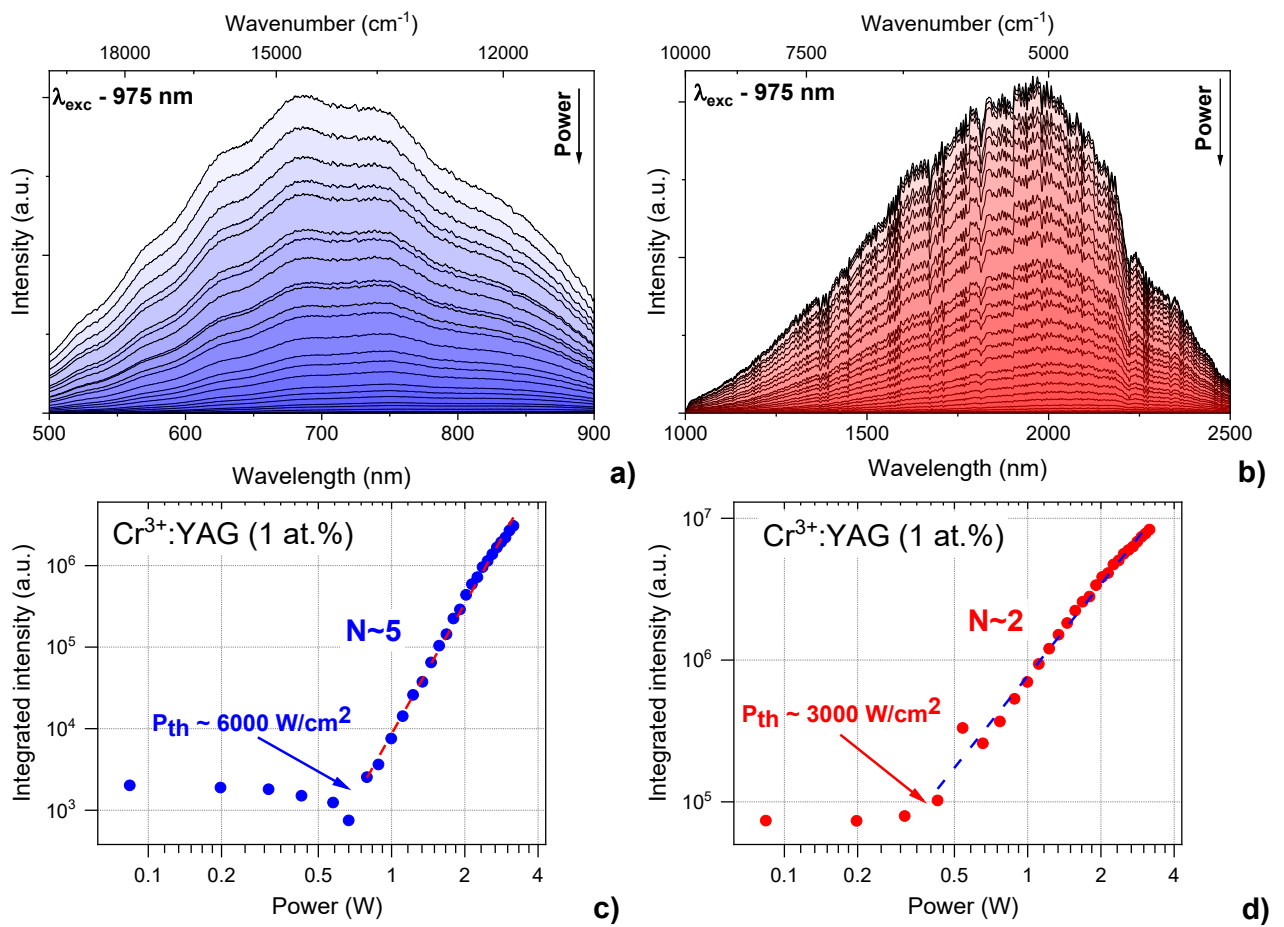

Fig. 6: LIWE spectra of Cr:YAG nanopowders for anti-Stokes (a) and the Stokes (b) emission. The power dependence of LIWE intensity for anti-Stokes (c) and Stokes (d) emission.

The main parameters of LIWE, namely power threshold and order of process (N parameter) were calculated from the energy dependence of the emission intensity. The main parameter of LIWE is the number of absorbed photons (N) involved in LIWE process. This value can be extracted from the power dependence of LIWE intensity. The intensity of white light emission increases exponentially with increasing the excitation power after exceeding the threshold. The number of photons N can be calculated from the slope of logarithmic plot of power dependence (Fig. 6(c,d)). It was found that the anti-Stokes part of the emission spectra can be generated after exceeding a threshold of 6000 W/cm$^2$, while the Stokes part has a threshold that is half as small - 3000 W/cm$^2$. The same threshold was for all concentrations of chromium ions.

It was found that an increase in the concentrations of $Cr^{3+}$ ions caused an increase in the N parameter for both anti-Stokes and Stokes parts of the emission spectra. An increase in the concentration of chromium leads to an increase in the number of absorbed photons for LIWE cycle (N-value). For Cr:YAG (0.1 at.%) nanopowders N = 5.0(5), while the samples with the highest chromium concentration (30 at.%) have N = 9.5(7), which is almost twice as high. It should be noted that pure YAG nanopowders have N = 5(1) that is close to N for Cr:YAG with the lowest concentration of $Cr^{3+}$ ions. The logarithmic plot of the concentration dependence of N shows a linear dependence for both the anti-Stokes and Stokes parts of the emission spectra with a slope of 0.8 and 0.4, respectively (Fig. 7). It should be noted that the sample with the chromium concentration of 1 at.% stands out from the general trend. The general trend predicts that the N parameter for the Cr1 sample should be 6.5 and 2.4 for the anti-Stokes and Stokes parts of emission spectra, whereas the actual values were 5 and 2.2, respectively. Interestingly, the Cr1 sample has the highest $Cr^{3+}$ photoluminescence intensity (Fig. 4(b)).

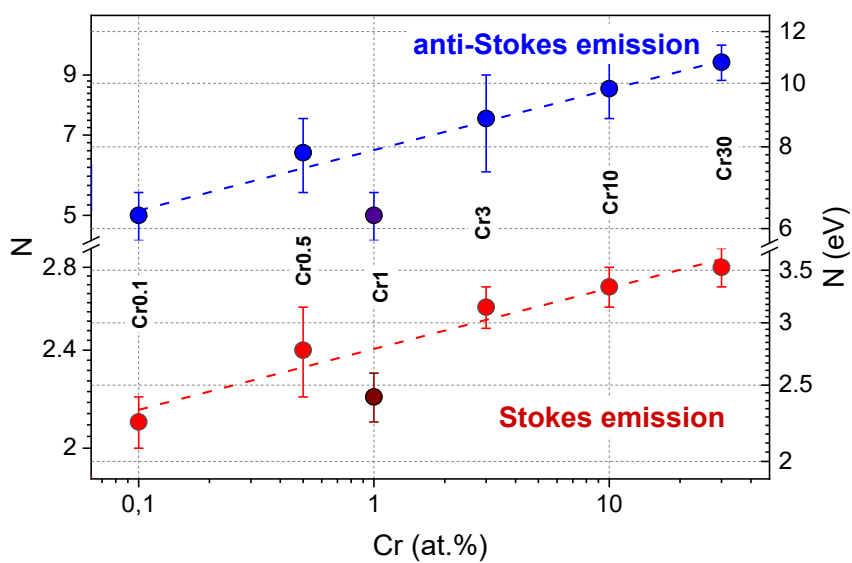

Fig. 7: Log-Log plot of concentration dependence of N value for anti-Stokes (blue) and Stokes (red) parts of LIWE spectra for Cr:YAG nanopowders.

## 4. Discussion

It has previously been proposed that the multiphoton ionization can be responsible for LIWE [2,8,9,13,26]. This assumption is based on the surface origin of LIWE meaning that white light is generated on the surface of the sample and do not penetrate inside the volume [2,8]. LIWE and multiphoton ionization have many similarities including threshold dependence, exponential growth, and saturation of emission intensity [2,10]. According to the multiphoton ionization model, irradiation of the sample with a focused laser beam leads to multiphoton absorption of the beam by $Cr^{3+}/Cr^{6+}$ mixed valence pair accompanied by electron transfer to conduction band [8]. The electrons can be captured by electron traps. If these electrons receive enough energy to exceed the ionization potential, ionization occurs. These electrons return to the sample and recombine producing photons and phonons [2]. The proposed model can generally explain the origin of LIWE, but it is far from perfect.

We assume that an increase in the N parameter is caused by an increase in the probability of non-radiative recombination for higher temperatures. The multiphoton ionization model cannot explain the growth of N with increasing concentration of doped ions. According to the proposed model, an increase in the concentration of doped ions should lead to an increase in the total emission intensity, while the ionization barrier should remain the same [9,10]. It should be noted that such a dependence of N parameter was found earlier for $Y_{2(1-x)}Nd_{2x}Si_2O_7$ nanocrystals [11], or even for an excitation pump power [27]. In both cases, increasing the excitation power or the concentration of doped ions leads to an increase of N. We suppose that this increase is caused by the negative influence of temperature on LIWE. It was previously shown that temperature increase negatively affects the LIWE properties leading to decrease in emission intensity [26] and increase in the rise and decay times [13]. Most likely, increasing temperature increase a probability of non-radiative recombination of excited electrons, so more photons need to be absorbed to produce LIWE. Moreover, a change in temperature affects the probability of NIR light absorption by $Cr^{3+}$ ions [24] which can be a possible source of non-radiative losses.

The present paper is only a first step towards understanding of the influence of non-radiative processes on LIWE properties, while many questions remain unsolved. It was previously shown that the chromium-doped phosphors are characterized by a high efficiency of light-to-heat conversion [28]. An increase in the concentration of chromium ions leads to an increase in the probability of non-radiative processes causing heating of the sample. This fact explains the results shown in Fig. 7, as an increase in the concentration of Cr ions leads to an increase in the number of photons required to produce LIWE emission due to an increase in the non-radiative transition probability in chromium ions. The only exception is the sample doped with 1 at% of chromium ions, for which N parameter is lower than for other nearby concentrations (Fig. 7). It is likely that the low value of N parameter is due to the smallest non-radiative losses (Fig. 5). This finding indicates that increasing efficiency of LIWE can be possible by reducing the probability of non-radiative transition processes. More in-depth study of the influence of other dopants on the properties of LIWE is required. Therefore, further work should include the study of the influence of the concentrations of various TM or/and RE ions on LIWE properties.

## 5. Conclusions

The synthesized Cr:YAG nanopowders are characterized by pure YAG phase and the presence of $Cr^{3+}$ and $Cr^{6+}$ ions. Concentration series of chromium-doped YAG nanopowders were synthesized by Pechini method. The microstructure of the samples was analyzed by XRD. The calculated lattice parameters were from 12.01 Å to 12.08 Å with an average grain size from 15 nm to 35 nm. Optical absorption spectra consist of five main absorption peaks centered at ~210 nm, 270 nm, 370 nm, 450 nm and 600 nm corresponding to absorption of color centers, $Cr^{3+}$, and $Cr^{6+}$ ions. The photoluminescence intensity and lifetimes of $Cr^{3+}$ ions increased with increasing chromium concentration reaching maximum for 1 at. % of $Cr^{3+}$ followed by subsequent decline.

It was found that LIWE threshold was the same for all samples, and N increased with chromium concentration. Under NIR laser irradiation in vacuum, Cr:YAG nanopowders emit white light covering entire Vis and NIR regions from 400 nm to at least 2600 nm. The anti-Stokes part of emission spectra is observed after exceeding a threshold of 6000 $W/cm^2$, while the Stokes part has a threshold of 3000 $W/cm^2$. For Cr:YAG nanopowders with chromium concentration of 0.1 at.% N parameter is 5.0(5), while for the samples with the highest concentration of chromium dopants (30 at.%) it is 9.5, i.e. almost twice higher. The logarithmic plot of the concentration dependence of N value shows a linear dependence for both anti-Stokes and Stokes part of the emission spectra with slopes of 0.8 and 0.4, respectively.

Multiphoton ionization in $Cr^{6+}/Cr^{3+}$ ion pair was used to explain the influence of concentration of chromium ions on N parameters. An increase in N parameter indicates an increase in the number of photons required to produce LIWE and can be caused by the negative influence of temperature during LIWE and the redistribution of electron traps with increasing chromium concentration. Most likely, increasing temperature stimulates non-radiative recombination of excited electrons, so more photons are required to be absorbed for LIWE.

**Acknowledgement**

This work was supported by Polish National Science Centre, grant: PRELUDIUM-18 2019/35/N/ST3/01018.